      \documentstyle[12pt,graphicx]{article}                
\begin{document}
\baselineskip 18pt
\def\today{\ifcase\month\or
 January\or February\or March\or April\or May\or June\or
 July\or August\or September\or October\or November\or December\fi
 \space\number\day, \number\year}
\def\thebibliography#1{\section*{References\markboth
 {References}{References}}\list
 {[\arabic{enumi}]}{\settowidth\labelwidth{[#1]}
 \leftmargin\labelwidth
 \advance\leftmargin\labelsep
 \usecounter{enumi}}
 \def\newblock{\hskip .11em plus .33em minus .07em}
 \sloppy
 \sfcode`\.=1000\relax}
\let\endthebibliography=\endlist
\def\lsim{\ ^<\llap{$_\sim$}\ }
\def\gsim{\ ^>\llap{$_\sim$}\ }
\def\r2{\sqrt 2}
\def\beq{\begin{equation}}
\def\eeq{\end{equation}}
\def\beqn{\begin{eqnarray}}
\def\eeqn{\end{eqnarray}}
\def\rmuu{\gamma^{\mu}}
\def\rmud{\gamma_{\mu}}
\def\PL{{1-\gamma_5\over 2}}
\def\PR{{1+\gamma_5\over 2}}
\def\sinW2{\sin^2\theta_W}
\def\AEM{\alpha_{EM}}
\def\mul{M_{\tilde{u} L}^2}
\def\mur{M_{\tilde{u} R}^2}
\def\mdl{M_{\tilde{d} L}^2}
\def\mdr{M_{\tilde{d} R}^2}
\def\mz2{M_{z}^2}
\def\c2b{\cos 2\beta}
\def\au{A_u}
\def\ad{A_d}
\def\cob{\cot \beta}
\def\v#1{v_#1}
\def\tb{\tan\beta}
\def\epem{$e^+e^-$}
\def\KK{$K^0$-$\bar{K^0}$}
\def\wi{\omega_i}
\def\xj{\chi_j}
\def\Wmu{W_\mu}
\def\Wnu{W_\nu}
\def\m#1{{\tilde m}_#1}
\def\mH{m_H}
\def\mw#1{{\tilde m}_{\omega #1}}
\def\mx#1{{\tilde m}_{\chi^{0}_#1}}
\def\mc#1{{\tilde m}_{\chi^{+}_#1}}
\def\mwi{{\tilde m}_{\omega i}}
\def\mxi{{\tilde m}_{\chi^{0}_i}}
\def\mci{{\tilde m}_{\chi^{+}_i}}
\def\mz{M_z}
\def\sw{\sin\theta_W}
\def\cw{\cos\theta_W}
\def\cb{\cos\beta}
\def\sb{\sin\beta}
\def\rwi{r_{\omega i}}
\def\rxj{r_{\chi j}}
\def\rfp{r_f'}
\def\Kik{K_{ik}}
\def\Fq2{F_{2}(q^2)}
\def\tw{\tan\theta_W}
\def\sec2w{sec^2\theta_W}

\begin{titlepage}
{\flushleft {NSF-ITP-99-129}\\
{NUB-TH-3206}}

\  \
\vskip 0.5 true cm
\begin{center}
{\large {\bf Large CP Phases and the Cancellation Mechanism in
 EDMs in SUSY, String and Brane Models }}\\

\vskip 0.5 true cm
\vspace{2cm}
\renewcommand{\thefootnote}
{\fnsymbol{footnote}}
 Tarek Ibrahim$^a$ and Pran Nath$^{b,c}$  
\vskip 0.5 true cm
\end{center}
\noindent
{a. Department of  Physics, Faculty of Science,
University of Alexandria,}\\
{ Alexandria, Egypt}\\ 
{b. Department of Physics, Northeastern University,
Boston, MA 02115-5000, USA } \\
{c. Institute for Theoretical Physics, University of California,
Santa Barbara, CA 93106-4030}\\
\vskip 1.0 true cm

\centerline{\bf Abstract}
\medskip
We show that EDMs obey a  simple  approximate 
scaling under the transformation  $m_0\rightarrow \lambda m_0,
 m_{\frac{1}{2}}\rightarrow \lambda m_{\frac{1}{2}}$ in the large 
 $\mu$ region when $\mu$ itself obeys the same scaling, ie., 
 $\mu \rightarrow \lambda \mu$. In the scaling region the
 knowledge of a single point in the MSSM parameter space 
 where the cancellation in the EDMs occur  allows one to generate 
 a trajectory in the $m_0-m_{\frac{1}{2}}$ plane where the cancellation
 mechanism holds and the EDMs are small. We illustrate these 
 results for MSSM with radiative  electro-weak 
 symmetry breaking constraints. We also discuss a  
 class of  D brane models based on Type IIB superstring 
 compactifications  which have
   non-universal phases in the gaugino mass sector and allow 
   large CP violating phases consistent with the EDM constraints
   through the cancellation mechanism. The scaling in these
   D brane models and in a heterotic string model is also
   discussed. 
\end{titlepage}

\section{Introduction}
Supersymmetric theories contain new sources of CP violation 
which arise from the phases of the soft SUSY breaking parameters which
are in general complex. The CP violating phases 
associated with the complex soft SUSY breaking parameters 
are  typically large, i.e. O(1), and pose a problem 
regarding the satisfaction of the current experimental limits 
on the neutron and on the electron EDM. For the neutron the 
current experimental limit is\cite{harris}
\beq
 |d_n|< 6.3\times 10^{-26} ecm
 \eeq
 and for the electron the limit is\cite{commins}
 \beq
 |d_e|< 4.3\times 10^{-27} ecm.
\eeq
Various remedies have been suggested in the literature to
overcome this problem. The first of these is the 
suggestion that the phases are small\cite{ellis,wein}
 $O(10^{-2})$.
However, small phases constitute a fine tuning and are 
thus undesirable. Another suggestion is that 
 the sparticle mass spectrum is heavy in the
several TeV range to suppress the EDMs\cite{na}. 
 A third possibility suggested is that 
there are internal cancellations among the various contributions 
to  the neutron and to the electron EDM leading to compatability 
with experiment with large phases and a SUSY spectrum that is
still within the reach of the accelerators\cite{in1}. 
There have been further 
developments\cite{fo,bgk,bartl,prs,everett,accomando} 
 and applications of this idea to explore the  effects of 
 large CP violating phases on
 dark matter 
 analyses\cite{ffo,cin,ks,gf,choi}, on $g_{\mu}-2$\cite{in3}, 
 and on other  
 low energy phenomena\cite{pilaftsis,pw,demir,kane,barger,more}. 
The focus of this paper is to show that in theories
where the higgs mixing parameter $\mu$ obeys the simple scaling
behavior as the rest of the SUSY masses the EDMs exhibit a simple 
scaling behavior under the simultaneous scaling on $m_0$ and
$m_{\frac{1}{2}}$. The scaling property of EDMs allows one to 
 promote a
single point in the SUSY parameter space where cancellations 
occur to a trajectory in the $m_0-m_{\frac{1}{2}}$ plane.
The scaling phenomena also  has implications
for the satisfaction of the EDM constraints in string 
and D brane models.
The outline of the paper is as follows:  In Sec.2 we discuss the
 scaling transformations and the properties of the relevant
 SUSY spectrum under scaling in the region of large $\mu$.
 In Sec.3 we  discuss the properties of the EDMs under scaling in this 
 region. In Sec.4 we discuss the algorithm for the
  satisfaction of the EDM constraints.
 We also investigate the parameter space where $\mu$ is large and
 show that in this region scaling can be used
 to generate trajectories in the
 $m_0-m_{\frac{1}{2}}$ plane where the cancellation mechanism holds. 
 The cancellation mechanism in 
  string models and D brane  models is  discussed in Sec.5.
  Conclusions are given in Sec.6.

\section{Scaling}
In this section we discuss the properties of the chargino 
and the neutralino mass eigen-values and eigen vectors
  under the scale transformation
\beq 
m_0\rightarrow \lambda m_0,~~~~m_{\frac{1}{2}}\rightarrow 
\lambda m_{\frac{1}{2}}
\eeq
In general the eigen-spectrum will have no simple property 
under this transformation since the chargino and the neutralino
mass matrices contain non-scaling parameters $M_W$ and $M_Z$.
 However, simple scaling properties emerge when
 $|\mu|>>M_Z$. 
In MSSM $\mu$ is an independent parameter and has no
scaling property under Eq.(3). However, in scenarios with 
 radiative breaking of the electro-weak symmetry $\mu$ 
is determined via one of the extrema equations by varying
the effective potential
\beq
\mu^2=\frac{1}{2}M_Z^2+\frac{\tilde m_{H1}^2 -\tilde m_{H1}^2 \tan^2\beta}
{\tan^2\beta-1}
\eeq
where $\tilde m_{H_i}^2=  m_{H_i}^2+\Sigma_i$ (i=1,2)
and where $\Sigma_i$ is the one loop correction to the Higgs 
mass. In the limit $|\mu|>>M_Z$,
$\mu^2$ becomes a homogeneous polynomial of degree 2 in $m_0$
and $m_{\frac{1}{2}}$, and thus under the transformation of Eq.(3) it has
the property 
\beq
\mu\rightarrow \lambda \mu
\eeq
From now on we shall consider the class of models where  Eq.(5) holds.
Next let us consider the chargino mass matrix  with the most
general set of phases
\beq
M_C=\left(\matrix{|\m2|e^{i\xi_2} & \r2 m_W  \sb e^{-i\chi_2}\cr
	\r2 m_W \cb e^{-i\chi_1}& |\mu| e^{i\theta_{\mu}}}
            \right)
\eeq
where our notation is as in Ref.\cite{in1}.
By the transformation
$ M_C=B_R M_C'B_L^{\dagger}$, 
where  $B_R=diag(e^{i\xi_2},e^{-i\chi_1})$ and 
$B_L=diag(1,e^{i(\chi_2+\xi_2)})$, the chargino mass matrix can
be written in the form 
\beq
M_C'=\left(\matrix{|\m2| & \r2 m_W  \sb \cr
	\r2 m_W \cb & |\mu| e^{i(\tilde{\theta})}}
            \right)
\eeq
where $\tilde{\theta}=\theta_{\mu}+\xi_2+\chi_1+\chi_2$.
We can diagonalize the matrix $M'_C$  by the biunitary transformation 
\beq
U_R^{'\dagger}M_C'U_L=diag(|\mc1| e^{i\gamma_1}, |\mc2| e^{i\gamma_2})
\eeq
In the limit of $|\mu|>(M_W, |\tilde {m_2}|)$
 the eigen-values of the chargino 
mass matrix are\cite{scaling}
\beq
\mc1\simeq |\mu|,~~ \mc2 \simeq |\m2|
\eeq
These relations were derived originally in the absence of CP violating
phases in the limit of large $\mu$ in supergravity models with radiatively 
induced breaking of the electro-weak symmetry.
 Here we find that
the relations continue to hold when CP violating phases are
included. 
 The matrices $U^{'}_R$ and $U_L$ in the large $\mu$ limit may
 be expanded as follows 
\beqn
U_{L}= 1+U^{(1)}_{L}(\frac{M_W}{|\mu|})
+U^{(2)}_L(\frac{M^2_{W}}{|\mu|^2})+....\nonumber\\
U^{'}_{R}= 1+U^{(1)}_{R}(\frac{M_W}{|\mu|})
+U^{(2)}_R(\frac{M^2_{W}}{|\mu|^2})+....
\eeqn
where  $U^{(1,2)}_{L,R}$ are scale independent matrices  and are given by
\beqn
U^{(1)}_L=\left(\matrix{0 & \r2 \cb e^{i\tilde{\theta}}\cr
        -\r2 \cb e^{-i\tilde{\theta}}& 0}
            \right)\nonumber\\
U^{(2)}_L=\left(\matrix{-\cos^{2}\beta & 0\cr
        0& -\cos^{2}\beta}
            \right)\nonumber\\
U^{(1)}_R=\left(\matrix{0 & \r2 \sb e^{-i\tilde{\theta}}\cr
        -\r2 \sb e^{i\tilde{\theta}}& 0}
            \right)\nonumber\\
U^{(2)}_R=\left(\matrix{-\sin^{2}\beta & 0\cr
        0& -\sin^{2}\beta}
	   \right)
\eeqn
By defining $U_R=U^{'}_R\times diag(e^{-i\gamma_1},e^{-i\gamma_2})$
 one can have
\beq
U_R^{\dagger}M_C'U_L=diag(|\mc1|,  |\mc2| )
\eeq
Thus to the leading order under the transformation of Eqs.(3) and (5) one has
\beq
|m_{\chi_i^{+}}|\rightarrow \lambda |m_{\chi_i^{+}}|,~~i=1,2
\eeq
and the relevant matrix elements of the EDMs will have
the following scale transformations:
\beqn
Im(U_{L2i}U^{*}_{R1i})\rightarrow \frac{1}{\lambda}
Im(U_{L2i}U^{*}_{R1i})\nonumber\\
Im(U_{L1i}U^{*}_{R2i})\rightarrow \frac{1}{\lambda}
Im(U_{L1i}U^{*}_{R2i})
\eeqn
We  discuss now the  
 neutralino mass matrix 
\beq
\left(\matrix{|\m1|e^{i\xi_1}
 & 0 & -\mz\sw\cb e^{-i\chi_1} & \mz\sw\sb e^{-i\chi_2} \cr
  0  & |\m2| e^{i\xi_2} & \mz\cw\cb e^{-i \chi_1}& -\mz\cw\sb e^{-i\chi_2} \cr
-\mz\sw\cb e^{-i \chi_1} & \mz\cw\cb e^{-i\chi_2} & 0 &
 -|\mu| e^{i\theta_{\mu}}\cr
\mz\sw\sb e^{-i \chi_1} & -\mz\cw\sb e^{-i \chi_2} 
& -|\mu| e^{i\theta_{\mu}} & 0}
			\right).
\eeq
We define the matrix X that diagonalizes $M_{\chi^0}$ so that
\beq 
X^T M_{\chi^0} X={\rm diag}(\mx1, \mx2, \mx3, \mx4)
\eeq
 In the limit $|\mu|>\{M_Z, |\tilde {m_1}|,|\tilde {m_2}|\}$ 
 the neutralino mass eigen-values  have the following 
 form
\beq
\mx1\simeq |\m1|, \mx2\simeq |\m2|, \mx3\simeq |\mu|,
\mx4\simeq |\mu|
\eeq
Again the scaling relations of Eq.(17) were originally derived in
the limit of large $\mu$ and no CP phases and our analysis shows that these 
relations continue to hold when large CP violating phases are
included. From Eq.(17) we find that in the large $\mu$ limit 
under the transformations of Eqs.(3) and (5) one has
\beq
m_{\chi_i^0}\rightarrow \lambda m_{\chi_i^0}~~(i=1-4)
\eeq
In the large $\mu$ limit the diagonalizing matrix X has the 
expansion 
\begin{eqnarray}
X=X^{(0)}+X^{(1)} (\frac{M_Z}{|\mu|})+O(\frac{M_Z^2}{|\mu|^2})
\end{eqnarray}
where $X^{(0),(1)}$ are scale independent matrices.
Now we discuss the behavior of the diagonalizing matrix $D$ of the
sfermion $(mass)^2$ matrix under the scaling transformations where
\beq
D^\dagger M_{\tilde{f}}^2 D={\rm diag}(M_{\tilde{f}1}^2,
              M_{\tilde{f}2}^2)
\eeq
For light flavors the scale transformations for the mass eigen states
are
\beq
M_{\tilde{fi}}\rightarrow \lambda M_{\tilde{fi}}~~(i=1-2)
\eeq
and the matrix elements of $D$ have the following transformations
under the scaling transformation of Eqs.(3) and (5):  
$D_{11},D_{22}\rightarrow D_{11},D_{22}$;  
$D_{12},D_{21}\rightarrow \frac{1}{\lambda}D_{12},D_{21}$. 
We note, however, that for the light flavors (the electron,
the up quark and the down quark) one has 
$|D_{12},D_{21}| < |D_{11},D_{22}|$. 
For the heavy flavors (i.e., the top and the bottom quarks) 
which are relevant to the
six dimensional purely gluonic operator, the behavior of the eigen values
and of the diagnalizing matrices are much more
complicated and will be discussed later.

\section{Scaling Properties of EDMs }
In the analysis below we shall use the notation of Refs.\cite{in1}.
However, we will make the notation explicit where necessary.  
The chargino contribution to the EDM of the up quark is given by
\beq
{d_{u-chargino}^{E}}/{e}=\frac{-\AEM}{4\pi\sinW2}\sum_{k=1}^{2}\sum_{i=1}^{2}
      {\rm Im}(\Gamma_{uik})
               \frac{\mci}{M_{\tilde{d}k}^2} [Q_{\tilde{d}}
                {\rm B}(\frac{\mci^2}{M_{\tilde{d}k}^2})+
        (Q_u-Q_{\tilde{d}}) {\rm A}(\frac{\mci^2}{M_{\tilde{d}k}^2})],
\eeq
\noindent
where $A(r)=(2(1-r)^{-2}(3-r+2lnr(1-r)^{-1})$ and
\beq
\Gamma_{uik}=\kappa_u V_{i2}^* D_{d1k} (U_{i1}^* D_{d1k}^*-
                \kappa_d U_{i2}^* D_{d2k}^*)
\eeq
and $\kappa_u=m_ue^{-i\chi_2}/\sqrt{2} M_W\sin\beta$. 
Because of the smallness of $m_u$, we can ignore the second part of
$\Gamma_{uik}$ and the bigger component of it would be
that of $k=1$ and it could be written in terms of $U_{L,R}$ as
\beq
\Gamma_{ui1}\simeq |\kappa_u| |D_{d11}|^2
 U_{L2i}U^{*}_{R1i}
\eeq
which under the scaling transformation behaves as
\beq
\Gamma_{ui1}\rightarrow \frac{1}{\lambda}\Gamma_{ui1}
\eeq
So the chargino component of the electric operator for the up quark
$d^{\chi+}_{u}$ has the scale transformation
\beq 
d^{\chi+}_{u}\rightarrow \frac{1}{\lambda^2} d^{\chi+}_{u}
\eeq
and the same transformation holds for the down quark and for the electron
\beq
d^{\chi+}_{d,e}\rightarrow \frac{1}{\lambda^2} d^{\chi+}_{d,e}
\eeq
The neutralino exchange contribution to a fermion is given by\cite{in1}

\beq
{d_{f-neutralino}^E}/{e}=\frac{\AEM}{4\pi\sinW2}\sum_{k=1}^{2}\sum_{i=1}^{4}
{\rm Im}(\eta_{fik})
               \frac{\mxi}{M_{\tilde{f}k}^2} Q_{\tilde{f}}
{\rm B}(\frac{\mxi^2}{M_{\tilde{f}k}^2})
\eeq 
where
\beqn
\eta_{fik} & &={(a_0 X_{1i} D_{f1k}^*
  + b_0 X_{2i}D_{f1k}^*-
     \kappa_{f} X_{bi} D_{f2k}^*)} {( c_0 X_{1i} D_{f2k}
     -\kappa_{f} X_{bi} D_{f1k})}
\eeqn
Here b=3(4) for $T_{3q}=-\frac{1}{2}(\frac{1}{2})$,
$a_0=-\r2 \tan\theta_W (Q_f-T_{3f})$, $b_0=-\r2 T_{3f}$, 
$c_0=\r2 \tan\theta_W Q_f$. $\kappa_u$ is defined following
Eq.(23) and $\kappa_{d,e}$ is given by 
 $\kappa_{d,e}=m_{d,e}e^{-i\chi_1}/\sqrt{2} M_W\cos\beta$.
 Because of the smallness of $\kappa_f$ one
can write $\eta_{fik}$ as
\beq
\eta_{fik}\simeq a_0c_0 X^{2}_{1i} D^{*}_{f1k}D_{f2k}
+b_0c_0 X_{1i}X_{2i} D^{*}_{f1k} D_{f2k}
\eeq
and by using the expansion of the matrix $X$ of Eq.(19) one can
write 
\beq
\eta_{fik}\simeq a_0c_0 X^{(0)2}_{1i} D^{*}_{f1k}D_{f2k}
+b_0c_0 X^{(0)}_{1i}X^{(0)}_{2i} D^{*}_{f1k} D_{f2k}
\eeq
Since the transformation for $ D^{*}_{f1k}D_{f2k}$ for $k=1,2$ is given by
\beq
 D^{*}_{f1k}D_{f2k} \rightarrow \frac{1}{\lambda} D^{*}_{f1k}D_{f2k}
\eeq
the neutralino contribution for the electric operator for both the quarks
and the leptons  behaves as:
\beq
{d_{f-neutralino}^E}\rightarrow \frac{1}{\lambda^2}{d_{f-neutralino}^E}
\eeq
Eqs.(27) and (33) imply that $d_e$  satisfies the 
scaling property
\beq
d_e\rightarrow \frac{1}{\lambda^2} d_e
\eeq
Next we discuss the EDM components for the quarks which 
contains the contributions from several operators, i.e., the 
electric dipole operator, the color dipole operator and 
 the purely gluonic dimension six operator. 
\beq
d_q=d_q^E+d_q^C+d_q^G
\eeq
 For the electric dipole the chargino and the neutralino contributions
 have already been discussed. For the gluino exchange contribution one has
\beq
{d_{q-gluino}^E}/{e}=\frac{-2 \alpha_{s}}{3 \pi}  m_{\tilde{g}}Q_{\tilde{q}} 
{\rm Im}(\Gamma_{q}^{11}) [\frac{1}{M_{\tilde{q}1}^2}
 {\rm B}(\frac{m_{\tilde{g}}^2}{M_{\tilde{q}1}^2}) -\frac{1}{M_{\tilde{q}2}^2}
{\rm B}(\frac{m_{\tilde{g}}^2}{M_{\tilde{q}2}^2})].
\eeq
where $\Gamma_{q}^{1k}=e^{-i\xi_3} D_{q2k} D_{q1k}^*$,
  $\Gamma_{q}^{12}=-\Gamma_{q}^{11}$ and 
\beq
{\rm Im}(\Gamma_{q}^{11})=\frac{m_q}{M_{\tilde{q}1}^2-M_{\tilde{q}2}^2}
        (m_0 |A_q| \sin (\alpha_q -\xi_3)+ |\mu| \sin 
        (\theta_{\mu}+\chi_1+\chi_2+\xi_3) |R_q|),
\eeq
In the $|\mu|/M_Z>>1$ limit  we find that
${\rm Im}(\Gamma_{q}^{11})$ scales as $1/\lambda$ under the scaling
of Eq.(3) and $d_{q-gluino}^E$ exhibits the same scaling behavior,
i.e.,$d_{q-gluino}^E\rightarrow \frac{1}{\lambda^2}d_{q-gluino}^E$. 

Next we consider the chromoelectric dipole moment $\tilde d^C$ contribution
to the quark EDM. It is given by 
\beq
 d_{q}^C= \frac{e}{4\pi}\tilde d_{q}^C \eta^c
\eeq  
  where $\eta^c$ is the renormalization group evolution of the
  chromo-electric operator from the electro-weak scale to the 
  hadronic scale and numerically $\eta^c\sim 3.3$. 
 Contributions to $\tilde d_q^C$ arise from the gluino, from 
 the chargino and from the neutralino exchanges and we reproduce 
 here the analytic expressions derived in Ref.\cite{in1}.  
\beq
\tilde d_{q-gluino}^C=\frac{g_s\alpha_s}{4\pi} \sum_{k=1}^{2}
     {\rm Im}(\Gamma_{q}^{1k}) \frac{m_{\tilde{g}}}{M_{\tilde{q}_k}^2}
      {\rm C}(\frac{m_{\tilde{g}}^2}{M_{\tilde{q}_k}^2}),
\eeq

\beq
\tilde d_{q-chargino}^C=\frac{-g^2 g_s}{16\pi^2}\sum_{k=1}^{2}\sum_{i=1}^{2}
      {\rm Im}(\Gamma_{qik})
               \frac{\mci}{M_{\tilde{q}k}^2}
                {\rm B}(\frac{\mci^2}{M_{\tilde{q}k}^2}),
\eeq

and 
\beq
\tilde d_{q-neutralino}^C=\frac{g_s g^2}{16\pi^2}\sum_{k=1}^{2}\sum_{i=1}^{4}
{\rm Im}(\eta_{qik})
               \frac{\mxi}{M_{\tilde{q}k}^2}
                {\rm B}(\frac{\mxi^2}{M_{\tilde{q}k}^2}),
\eeq
where our notation is as in Ref.\cite{in1}.  
The expansion of these contributions
 in the limit $|\mu|/M_Z|>>1$ following the same procedure
 as for the electric dipole case shows that in this limit
  $\tilde d^C$  again shows the scaling behavior 
  $\tilde d^C \rightarrow \frac{1}{\lambda^2} \tilde d^C$ 
  under the transformations of
  Eqs.(3) and (5). 
  Finally, we consider the contribution of the purely gluonic 
  dimension six operator.   
  It is given by 
  \beq
  d_q^G=\frac{eM}{4\pi}\tilde d_q^G\eta^G
 \eeq
where $\eta^G$ is the renormalization group evolution of 
the purely gluonic dimension six operator from the 
electro-weak scale down to the hadronic scale ($\eta^G\simeq 3.3$)
and M is the chiral symmetry breaking scale (M$\simeq$1.19 GeV) 
and $\tilde d_q^G$ is given by\cite{wein} 
\beq
\tilde d_q^G=-3\alpha_s(\frac{g_s}{4\pi m_{\tilde g}})^3
(m_t(z_1^t-z_2^t)Im(\Gamma^{12}_t)H(z_1^t,z_2^t,z_t)
+m_b(z_1^b-z_2^b)Im(\Gamma^{12}_b)H(z_1^b,z_2^b,z_b))
\eeq
\noindent 
where
\beq
\Gamma_q^{1k}=e^{-i\xi_3}D_{q2k}D_{q1k}^*,
z^q_{\alpha}=(\frac{M_{\tilde{q}\alpha}}{m_{\tilde{g}}})^2,
z_q=(\frac{m_q}{m_{\tilde{g}}})^2\nonumber\\
\eeq
The behavior of $\Gamma^{12}_t$, $z^q_{\alpha}$ and $z_q$ under the scaling
transformation is a complicated one because of the largeness of the quark
masses involved and even if we were in a region where one can ignore these
masses compared to the other mass scales in the problem one finds that the
behavior of $d^G$ is different from that of the other components i.e.
$d^G\rightarrow \frac{1}{\lambda^4}d^G$.
Thus the scaling property of $d_q$ will be more complicated. 
However, as $\lambda$ gets large the contribution of $d_q^G$ will
fall off faster than the contribution of $d_q^E$ and $d_q^C$ and
in this case one will have the scaling 
$d_q\rightarrow \frac{1}{\lambda^2} d_q$ and so 
also the neutron edm $d_n$ will behave as
\beq
d_n\rightarrow \frac{1}{\lambda^2} d_n
\eeq
We note, however, that the question of how soon the scaling sets in
as we scale in $\lambda$ depends on the part of the parameter space one
is in.

\section{Satisfaction of EDM Constraints} 
In the work of Ref.\cite{in1} it was shown that the quark and the
lepton EDMs in general depend on ten independent phases which 
were classified there providing one with considerable freedom
for the satisfaction of the EDM constraints. Numerical analyses
 show the existence of significant regions of the 
parameter space where the cancellation mechanism holds. 
 We describe below a straightforward technique 
for accomplishing the satisfaction of the EDM constraints.
These techniques are already well understood and we codify 
them here for the benefit of the reader. 
For the case of the electron one finds that the
chargino component of the electron is independent of $\xi_1$ and
the electron EDM as a whole is independent of $\xi_3$. Thus the 
algorithm to discover a point of simultaneous cancellation for the 
electron EDM and for the neutron EMD is a straightforward one. For a given 
set of parameters except $\xi_1$ we start varying
 $\xi_1$ till we reach the cancellation for the electron EDM 
since only one of its components (the neutralino) is affected by that 
parameter. Once the electric dipole moment constraint on the 
electron is satisfied we vary $\xi_3$ which affects 
only the neutron edm keeping all other parameters fixed. By using this
simple algorithm one can generate any number of simultaneous
cancellations. In the numerical analysis of the EDMs we also
take into account the two loop diagrams of the type discussed
in Ref.\cite{darwin}. However, we find that in the small $\tan\beta$
region these diagrams do not make any substantial contributions to the
EDMs.

 We discuss now  the lepton and the neutron EDMs in the region
 where the scaling relation on the lepton and the neutron
 EDMs of Eqs.(34) and (45) hold.  Suppose 
 we have a point in the parameter space where the lepton and
 the quark EDMs vanish, i.e., $d_e=0, d_q=0$.
The interesting observation is that this cancellation 
constraint is preserved under
scaling provided one is in the scaling region, i.e., Eqs.(34) and
 (45) hold. Thus given a point in the parameter space where
 cancellations occur one can generate a trajectory in the 
 $m_0- m_{\frac{1}{2}}$ plane by a simple scaling of 
 $m_0$ and $m_{\frac{1}{2}}$ using Eqs.(3) and (5). In practice
 the cancellation is not designed to be perfect and the scaling
 properties of $d_e$ given by Eq.(34) and of $d_n$  given by Eq.(45)
 are only approximate. 
 Thus under the scaling transformation some minor adjustment 
 of the other parameters will in general be necessary. The length
of the trajectory  depends on the part of the parameter space one is
in. For some cases it is found that the trajectory can be 
long enough to cover the range of the parameter space consistent
with naturalness. An example of this phenomenon is shown in
 Fig.1  where five trajectories are generated, and where 
 each trajectory is generated from a single cancellation point
 for low values of $m_0$ and $m_{\frac{1}{2}}$ by
 simple scaling. We notice, however, that there is an empty region in 
trajectory 5 where the cancellation under scaling does not hold.
However, we have checked that with  a
very minor adjustments in the values of the other parameters we can 
restore the cancellation. 
 Thus each of the trajectories satisfy
 the  EDM constraints with the values of $A_0$, $\tan\beta$, and
 phase angles fixed  as we move along the trajectory. 
 As we move on the trajectory to the higher mass regions we have
 a  natural  suppression besides the cancellation suppression. 
 However, the cancellation is still  necessary except for the extreme
 ends of each trajectory.  
 In Fig.2 we exhibit the 
 EDM of the neutron corresponding to the five trajectories of
 Fig.1. We find that all the trajectories are consistent with
 the current experimental constraint on the neutron EDM.
 In Fig.3 we plot the EDM of the electron   
  corresponding to the five trajectories of
 Fig.1. Again we find that all the trajectories are consistent with
 the current experimental constraint on the electron EDM.

	In summary a convenient procedure for generating  a
	trajectory in the $m_0-m_{\frac{1}{2}}$ plane where
	cancellations  of the EDMs occur, consists of finding
	a single point in the MSSM parameter space with low
	values of $m_0$ and $m_{\frac{1}{2}}$ under the constraint
	of the radiative breaking of the electro-weak symmetry 
	using the algorithm described in the beginning of this
	section where the cancellation in EDMs of the electron and 
	of the neutron occur consistent with Eqs.(1) and (2).
	One then computes the EDMs using Eqs.(3) and (5) for
	$\lambda >1$ and typically one finds that the EDM
	constraints are maintained with only minor adjustment
	of other parameters. The onset of the scaling behavior 
	itself will depend on the values of the other MSSM 
	parameters. We emphasize that in some cases the
	subleading terms in the scaling law may be significant
	and could generate new cancellations if they change
	sign as we scale upward in $\lambda$. While such points
	violate the scaling law, they are nonetheless acceptable
	since there is an even greater satisfaction of the EDM 
	constraints for this case.
	In Figs. 4 and 5 we plot $log_{10} \lambda^2 |d_{e,n}|$ as
	a function of $m_{1/2}$ 
	and we see support of the scaling idea here.
	 It is important to keep in mind that the
	method we outlined here is only an approximation and should
	be used keeping that in mind. The method  would
	work best if one is in the scaling region or close to
	it. Certainly it should be of relevance in exploring 
	at least a part of the parameter space where these
	conditions are met.

\section{String and Brane Models and EDM cancellations}
We discuss now CP violation and  cancellations in EDMs for 
the case of string and brane models.
Recently,  the progress
in string dualities has led to the formulation of a new  class of
models based on M theory compacitified on $CY\times S^1/Z_2$ and
models in the framework of Type IIB orientifolds. We shall focus
here on Type IIB orientifold models which have received
significant attention recently\cite{berkooz}. Specifically we shall consider 
models with compactification of 
the Type IIB theory on a six-torus $T^6=T^2\times T^2\times T^2$
of the type discussed in Ref.\cite{ibanez}.
In scenarios of this type
as in SUSY models and other string models additional sources 
of CP violation can arise through the breaking of 
supersymmetry. The mechanism of breaking of supersymmetery here
is not fully understood. However, one can
still make some progress by phenomenologically parametrizing how
supersymmetry breaks.
An efficient way of doing so is in terms of 
the VEVs of the dilaton field (S) and of the moduli fields $T_i$ and for 
the case when the vacuum energy is set to zero one has that 
F type supersymmetry breaking may be parametrized by\cite{ibanez} 
\begin{eqnarray}
F^S=\sqrt{3} m_{\frac{3}{2}} (S+S^*)\sin\theta e^{-i\gamma_S}\nonumber\\
F^i=\sqrt{3} m_{\frac{3}{2}} (T+T^*)\cos\theta \Theta_i e^{-i\gamma_i}
\end{eqnarray}
where $\theta$, $\Theta_i$ parametrize the Goldstino direction in the
 S, $T_i$ field space and $\gamma_S$ and $\gamma_i$ are the $F^S$ and
$F^i$ phases, and $\Theta^2_1+\Theta^2_2+\Theta^2_3=1$.
 The Type IIB compactified models of the  type mentioned above
  contain 9 branes,
$7_i$ (i=1,2,3) branes, $5_i$ (i=1,2,3) branes and 3 branes.
 N=1 supersymmetry constraints require that not all the branes can
 simultaneously be present, and thus one can have either 9 branes and $5_i$
 branes or $7_i$ branes and 3 branes. Recently the work of 
 Ref.\cite{everett} investigated the EMD constraints on models based
 on $5_i$ (i=1,2) branes which belong to the general class of models
 discussed in Ref.\cite{ibanez}. It was shown and that 
 this model exhibits non-universalities in the phases of the  gaugino
 masses and  that cancellations in the EDMs arise
 and one can achieve satisfaction of the EDM
 constraints consistent with experiment\cite{everett,accomando}.
 Our own analysis of this model further confirms the existence of 
 the cancellations for the EDMs in the parameter space of this model.

  We discuss here the models based  on 
 9 branes and one from the set of $5_i$ branes  which we choose to
 be $5_1$ where the Standard Model gauge group is distributed 
 between the two branes. 
 Like the models  based on $5_i$ (i=1,2) branes, 
 these models also contain non-universalities of the gaugino
 phases due to different gauge kinetic energy functions associated
 with 9 branes and $5_i$ branes, i.e., $f_9=S$, and $f_{5_i}=T_i$.
  However, the nature of the soft SUSY breaking is 
 different in these models from the ones  based on $5_i$ branes.
 Thus it is interesting to investigate the question of 
 large CP violating phases and of cancellations in the EDMs in 
 this type of models. 
In models where more than one type of branes are involved the 
unification of the gauge couplings  requires
fine tuning.  For the  case
of models based on the 9 brane and the $5_1$ brane the unification
of gauge couplings is more difficult than in the case when the 
gauge groups are embedded on two different same type branes. 
A full discussion of  this topic is outside the scope of this work.
However, we wish to note that  contributions  from extra matter and
twisted moduli\cite{ibanez} could be important in a realistic 
analysis of the gauge coupling unification in this case. 
 For the purpose of the analysis we shall simply assume that the
 unification does occur at the usual scale of $\sim 10^{16}$ GeV.
We emphasize that the issue of
cancellations in the EDMs is largely independent of the issue
of the gauge coupling unification and thus the conclusions of our
analysis are largely independent of this issue.

Below we consider the following two ways to embed the Standard Model 
gauge group on the 9 branes and $5_1$ branes.\\
\noindent
Case I:\\
 Here we consider the possibility that the $SU(3)_C\times U(1)_Y$
is associated with the 9 brane and the $SU(2)_L$  is 
associated with the $5_1$ brane. Further we assume that the 
$SU(2)_R$ singlet states are associated with the nine-brane
sector, while the $SU(2)_L$ doublet states arise from the 
intersection of 9-brane and $5_1$-brane sector as in Case I. In this model
we find using the general formulae of Ref.\cite{ibanez} the following
results: the $SU(2)_R$ singlets have the common mass
$m_9$ and the $SU(2)_L$ doublets have the common mass 
$m_{95_1}$ where    
\begin{equation}
m_9^2=m^2_{\frac{3}{2}}(1-3\cos^2\theta \Theta_1^2)
\end{equation}
\begin{equation}
m_{95_1}^2=m^2_{\frac{3}{2}}(1-\frac{3}{2}\cos^2\theta (1-
\Theta^2_1)).
\end{equation}
The $SU(3)$, $SU(2)$ and $U(1)$ gaugino masses $\tilde m_i$ (i=1,2,3)
 are given by 
\begin{eqnarray}
\tilde {m}_1=\sqrt{3} m_\frac{3}{2}\sin\theta e^{-i\gamma_S}
=\tilde m_3=-A_0,\nonumber\\
\tilde {m}_{2}=\sqrt{3} m_\frac{3}{2}\cos\theta 
\Theta_1 e^{-i\gamma_1}
\end{eqnarray}
In the analysis of the EDMs we shall treat the phase of
 $\mu$ to be a free parameter and the magnitude of $\mu$ is 
 determined by the radiative breaking of the electro-weak symmetry.
 In order to avoid tachyons we  impose the constraint 
 $cos^2\theta \Theta_1^2 < 1/3$. 
In Fig.6 we exhibit the cancellation phenomenon for the
EDMs for this case in the presence of large CP violating 
phases.\\
\noindent
Case II:\\
 The second possibility is that the $SU(3)_C\times U(1)_Y$
is associated with the $5_1$ brane and the $SU(2)_L$  is 
associated with the $9$ brane. 
Regarding the  matter fields we assume that the 
$SU(2)_R$ singlet states are associated with the $5_1$
sector, while the $SU(2)_L$ doublet states arise from the 
intersection of the $5_1$-brane and the 9-brane sector. 
Although this case is T dual to 
Case I the pattern of soft masses is different
after the breaking  of supersymmetry in the two cases.
Thus after SUSY breaking one finds here that the $SU(2)_R$ singlet 
masses have the 
common mass $m_{5_1}$ and the $SU(2)_L$ doublet masses have the 
common mass $m_{95_1}$ where    
\begin{equation}
m_{5_1}^2=m^2_{\frac{3}{2}}(1-3\sin^2\theta)
\end{equation}
\begin{equation}
m_{95_1}^2=m^2_{\frac{3}{2}}(1-\frac{3}{2}\cos^2\theta (1-
\Theta^2_1))
\end{equation}
 while the $SU(3)$, $SU(2)$ and $U(1)$ gaugino masses are given by 
\begin{eqnarray}
\tilde{m}_{1}=\sqrt{3} m_\frac{3}{2}\cos\theta \Theta_1 
e^{-i\gamma_1}=\tilde m_3 =-A_0,\nonumber\\
\tilde{m}_{2}=\sqrt{3} m_\frac{3}{2}\sin\theta e^{-i\gamma_S}
\end{eqnarray}
To guarantee that there  are no  tachyons we impose the  constraint
$sin^2\theta < 1/3$.  
We note that although one can go from Case I to Case II and vice versa by
the transformation $sin\theta \leftarrow \rightarrow  cos\theta  \Theta_1$
and $\gamma_S\leftarrow \rightarrow \gamma_1$, these cases are 
physically different. This is so because  once  $\theta$ and 
$\Theta_1$ which parametrize the goldstino direction in the dilaton and
the moduli VEV space are frozen,  these cases will lead to different 
sparticle masses and  will have physically distinct experimental
consequences.  Of course it is possible to view  the two cases as
part of a single case  with a larger parameter space but we
prefer to treat them as  distinct on physical grounds. Again as in 
  Case I we treat the phase of $\mu$ to be a free parameter and use the  
 radiative breaking of the electro-weak symmetry to determine the
 magnitude of $\mu$.
An exhibition of  the cancellation in 
EDMs for this case in the presence of large CP violating 
phases is given in Fig.7.

 	An interesting aspect of string models is that under the
	single scaling
	
	\begin{equation}
m_{\frac{3}{2}}\rightarrow \lambda m_{\frac{3}{2}} 
 \end{equation}
	one has  $F^S\rightarrow \lambda F^S$ and  
	$F^i\rightarrow \lambda F^i$ 
	and thus all the soft  SUSY breaking parameters will 
	have that scaling. 
 	We examine now the scaling phenomenon for the two 
 	models considered above.
	For this purpose it is useful to define 
	$\lambda =\frac{m_{3/2}}{m^0_{3/2}}$ where $m_{3/2}$ 
         is the running value and $ m^0_{3/2}$
	is $m_{3/2}$ at the extreme left.
In Fig.8 we exhibit the result of the extrapolations for
 $log_{10}\lambda^2 |d_{e,n}|$ as a function of 
$m_{\frac{3}{2}}$ starting from  a single point of cancellation
at the far left.   One finds that as $m_{\frac{3}{2}}$ increases
the scaling is obeyed here to a good approximation.
	For comparison we also consider  a heterotic  string model.  
	The cancellation for the EDMs for the type O-II model 
	of Ref.\cite{brig} was discussed in Ref.\cite{everett}. 
	We discuss here the scaling property.The 
	soft SUSY breaking sector of this theory is 
	parameterized by\cite{brig} 
\begin{equation}
 m^2_0=\epsilon' (-\delta_{GS})m^2_{\frac{3}{2}}
 \end{equation}

\begin{equation}
\tilde m_i=\sqrt{3} m_{\frac{3}{2}}(\sin\theta e^{-i\alpha_S}
-\gamma_i \epsilon \cos\theta e^{-i\alpha_T})
\end{equation}
where
$\gamma_1=-\frac{33}{5}+\delta_{GS}, \gamma_2=-1+\delta_{GS},
\gamma_3=3+\delta_{GS}$ 
and 
\begin{equation}
A_0=-\sqrt{3}m_{\frac{3}{2}}\sin\theta e^{-i\alpha_S}
\end{equation}
The parameter $\delta_{GS}$ is  fixed  by the constraint of 
anomaly cancellation in a given orbifold  model. 
The parameter $\mu$ and its phase are again treated as  independent
parameters.
In Fig.9 we exhibit the result of the extrapolations for
$log_{10}\lambda^2 |d_{e,n}|$  as a function of 
$log_{10}m_{\frac{3}{2}}$ starting from a single point of cancellation
at  the far left. We find that scaling is obeyed for two of the three 
cases exhibited in Fig.9 over the entire range of $m_{\frac{3}{2}}$ considered.
For the third case the initial part of the curves is in the 
non-scaling region and a new cancellation appears which,
however, further reduces the EDM for this case maintaining 
consistency with the experimental EDM constraints. Eventually
of course 
scaling seems to set in for this case as $m_{3/2}$ becomes larger.
This third example is an interesting illustration of the approximate
nature of the scaling analysis and of subleading non-scaling corrections.
 Since the cancellation is a rather delicate phenomenon these
subleading terms can trigger a further cancellation which
would lead to a breakdown of scaling. 
However, the EDM constraints are satisfied 
 even more so in this case.

\section{Conclusion}
In this paper we discussed an algorithm for 
generating cancellations for the EDMs of the leptons and of
the quarks within the framework of MSSM. We  showed that
in theories where the $\mu$ parameter obeys the simple scaling
behavior of Eq.(5) under the scaling of Eq.(3),  the lepton
and the quark EDMs show a simple scaling property 
in the $m_0-m_{\frac{1}{2}}$ plane  in the large $\mu$ region. 
Thus in this region the cancellation constraint 
on the electron and on the quark EDMs is essentially maintained 
 under scaling. Thus given a single point in the SUSY parameter
space in the large $\mu$ region where cancellations occur
one can generate a trajectory in the $m_0-m_{\frac{1}{2}}$
plane where cancellations are maintained by the use of scaling 
with only minor adjustments in other parameters.
We emphasize that for low values of $m_0$ and $m_{\frac{1}{2}}$
some adjustment of the
parameters to satisfy the EDM constraints 
will in general be needed to compensate for the fact
that one is in the non-scaling region.  
  We also discussed a class of Type IIB string models with
  9 branes and $5_1$ branes which have non-universal 
  phases for the gaugino masses. We showed that such  
  models can  have large CP violating phases consistent  with 
  cancellations to guarantee the satisfaction of the EDM 
  constraints. We also exhibited the existence of scaling 
  in these models as well as in a heterotic string model.
 The simple algorithm described above with the caveats already
 discussed opens another  window  
 for the exploration of  the SUSY parameter space with large 
 CP phases and a relatively light SUSY particle spectrum.
Finally as already pointed out in the second paper of Ref.\cite{in1}
 the cancellation hypothesis is an experimentally testable idea,
 i.e., that  with soft SUSY phases $O(1-10^{-1})$ and with the
 SUSY spectrum within the naturalness limits of O(1) TeV, the
 EDM of the electron and of the neutron should become visible 
 with an order of magnitude improvement in the experimental 
 EDM measurements. We further point out here that this observation
   is generic and should cover a range of models whether  SUSY,
   string or brane. Such an order of magnitude improvement in 
   experiment should be  possible in the near future.

\noindent
{\bf Acknowledgements}\\ 
This work was done in part during the period when one of us (P.N.)
was participating in the ITP Program "Supersymmetry Gauge Dynamics
and  String  Theory". 
We wish to thank Zurab Kakushadze, Gary Shiu and especially 
Carlos Munoz for 
helpful discussions. This research was supported in part by NSF grants 
PHY-9901057 and PHY94-07194. \\

\newpage

\section{Figure Captions}
Fig.1. The trajectories in the $m_0-m_{\frac{1}{2}}$ plane 
generated by scaling where cancellations occur in the SUSY EDMs 
consistent with the EDM constraints. (1)$|A_0|=6.5$, $\theta_{\mu}=2.92$
, $\alpha_{A0}=-.4$, $\tan\beta=4$, $\xi_1=0$, $\xi_2=.2$ and  $\xi_3=.06$.  
 (2)$|A_0|=2.9$, $\theta_{\mu}=3.02$,
$\alpha_{A0}=.5$, $\tan\beta=2.6$, $\xi_1=.19$, $\xi_2=.19$ and  $\xi_3=.41$.
 (3)$|A_0|=5.5$, $\theta_{\mu}=3.006$,
$\alpha_{A0}=-.1$, $\tan\beta=3.5$, $\xi_1=.105$, $\xi_2=.105$ and  $\xi_3=.15$.
(4)$|A_0|=4.4$, $\theta_{\mu}=3.02$,
$\alpha_{A0}=-.6$, $\tan\beta=7$, $\xi_1=0$, $\xi_2=.1$ and 
$\xi_3=-.065$.
(5)$|A_0|=3.2$, $\theta_{\mu}=2.8$,
$\alpha_{A0}=-.4$, $\tan\beta=5$, $\xi_1=.31$, $\xi_2=.3$ and 
$\xi_3=.32$.

\noindent    
Fig.2. Plot of $log_{10}|d_e|$ of the electron edm vs $m_{\frac{1}{2}}$ 
for the five cases of Fig.1. 

\noindent
Fig.3. Plot of $log_{10}|d_n|$ of the neutron edm vs $m_{\frac{1}{2}}$
for the five cases of Fig.1.

\noindent

Fig.4. Plot of $log_{10}|\lambda^2 d_e|$ vs $m_{\frac{1}{2}}$ for the 
points in Fig. 2.

\noindent

Fig.5.  Plot of $log_{10}|\lambda^2 d_n|$ vs $m_{\frac{1}{2}}$
for the points in Fig.3.
\noindent

Fig.6. Plot of $log_{10}|d_{e,n}|$ vs $\theta_{\mu}$ for Model I based on
 9 branes and $5_1$ branes  exhibiting
 the cancellation of the EDMs for the electron and for the
 neutron for the case when $m_{3/2}=250$ GeV, $\theta=1$,
  $\tan\beta$=5, $\gamma_S$=0.295, 
 $\gamma_1$=0.409, $\theta_1$=0.64. The solid line is for the electron case
and the dashed one is for the neutron.

\noindent
Fig.7. Plot of $log_{10}|d_{e,n}|$ vs $\theta_{\mu}$ for Model II based on
 9 branes and $5_1$ branes  exhibiting
 the cancellation of the EDMs for the electron and for the
 neutron for the case when $m_{3/2}=500$ GeV, $\theta=0.3$,
  $\tan\beta$=5, $\gamma_S$=0.3, 
 $\gamma_1$=0.4, $\theta_1$=0.9. The solid line is for the electron case
and the dashed one is for the neutron.

\noindent
Fig.8. Plot of $log_{10}\lambda^2|d_{e,n}|$ vs $m_{3/2}$ using 
one cancellation point (at far left) for each of the cases in Figs. 6 and 7 
and scaling in $m_{3/2}$.
 $\theta_{\mu}$ for each curve is fixed at  the initial
point  to satisfy the experimental limits of edms by cancellation.

\noindent
Fig.9. Plot of $log_{10}\lambda^2|d_{e,n}|$ vs $log_{10}m_{3/2}$ 
for the heterotic string model discussed in the text using 
one cancellation point (at far left) for each of the three cases.
The parameters for the cases considered are: 
(1). $m_{3/2}$=1050 GeV, $\theta$=0.06, $\theta_{\mu}$=0.3,
$\tan\beta$=6, $\alpha_S$=0.15, $\alpha_T$=0.4,
$\delta_{GS}=-10$, $\epsilon$=0.006, $\epsilon'$=0.001,
(2)   $m_{3/2}$=340 GeV, $\theta$=0.6, $\theta_{\mu}$=0.3,
$\tan\beta$=3, $\alpha_S$=0.25, $\alpha_T$=0.37,
$\delta_{GS}=-4$, $\epsilon$=0.001, $\epsilon'$=0.05,
(3) $m_{3/2}$=3 TeV, $\theta$=0.05, $\theta_{\mu}$=0.5,
$\tan\beta$=8, $\alpha_S$=0.39, $\alpha_T$=0.59,
$\delta_{GS}=-8$, $\epsilon$=0.004, $\epsilon'$=0.0012.


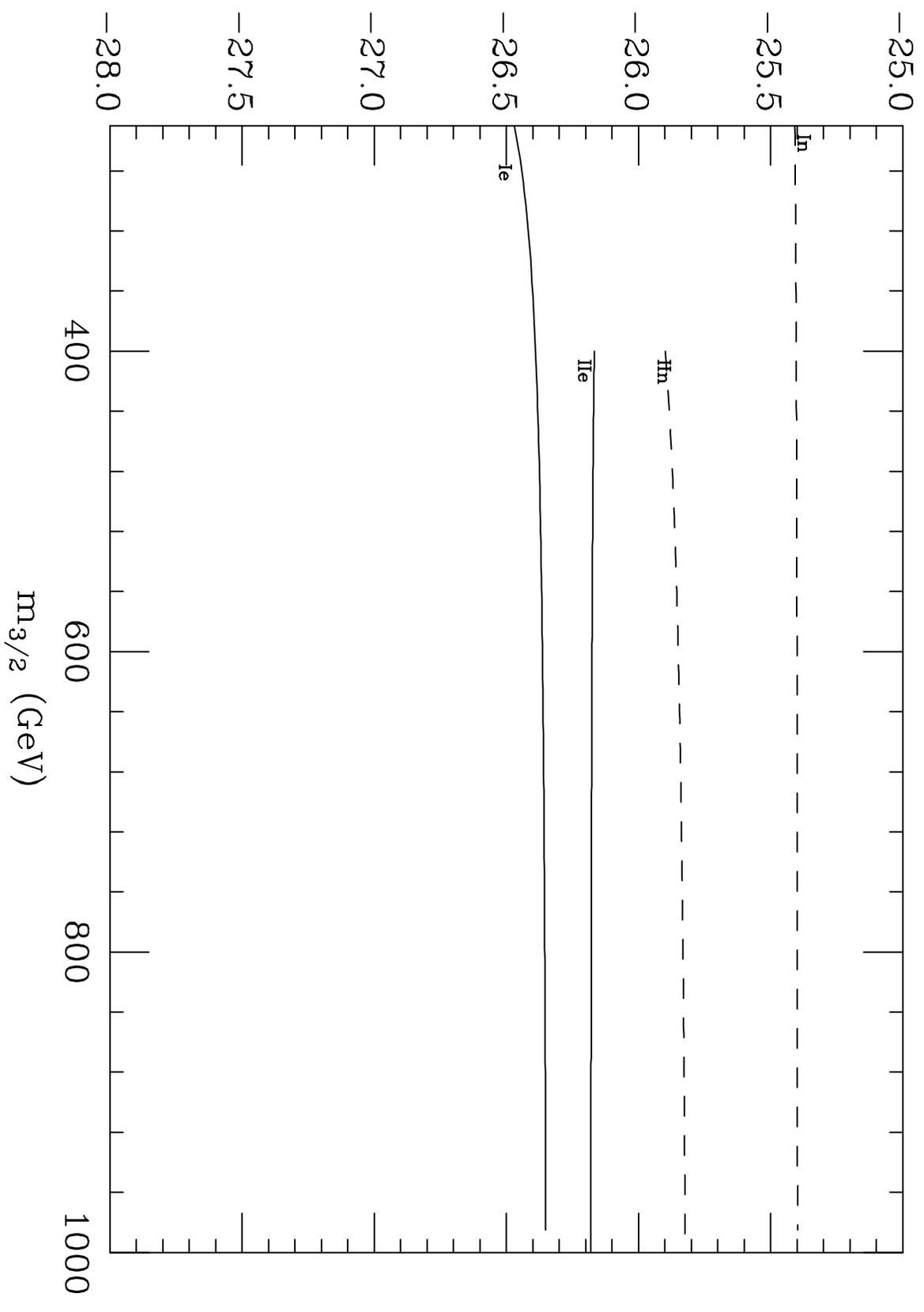
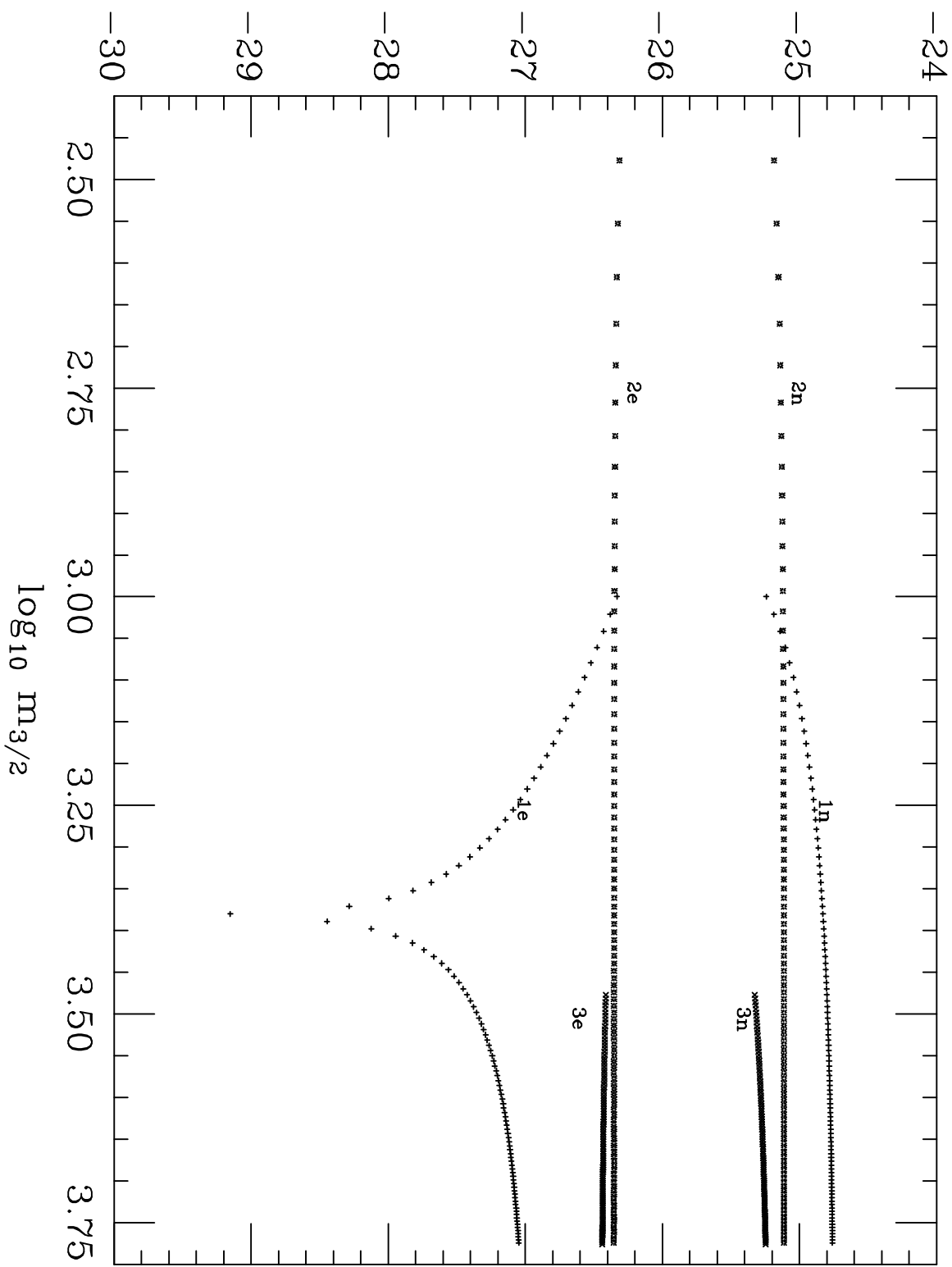
\end{document}